# Main parameters of SppC-based "linac-ring $eA$" and "ring-ring $\mu A$" colliders

Bora Ketenoglu

Department of Engineering Physics, Ankara University, 06100 Tandogan, Ankara, Turkey
*bketen@eng.ankara.edu.tr*

**Abstract**

Concerning future lepton-nucleus colliders, International Linear Collider (ILC) and Plasma Wake Field Accelerator-Linear Collider (PWFA-LC) electrons in order of 0.5 TeV and 5 TeV, respectively, colliding with Super proton-proton Collider (SppC)'s 3075 TeV lead ions, are considered as "linac-ring $eA$" options. In addition, 1.5 TeV option of the Muon Collider (MC) vs 3075 TeV $^{208}Pb^{82+}$ ions of the SppC is also taken into account as a "ring-ring $\mu A$" collider. Luminosity values of the SppC-based $eA$ and $\mu A$ colliders are estimated. 3075 TeV lead parameters for 100 km-circumference option of the SppC are taken into account to optimize luminosities of electron-nucleus and muon-nucleus collisions, keeping beam-beam effects and disruption in mind. It is shown that luminosities of order of $10^{30}$ cm$^{-2}$s$^{-1}$ can be achieved by moderate upgrades of lepton and nucleus beam parameters.

## 1. Introduction

Lepton-hadron collisions played crucial role in our understanding of inner structure of matter. For instance, proton form-factors and quark-parton model have been established by fixed target experiments using GeV-energy electron beams, first and sole $ep$ collider HERA has provided important information on proton structure as well as parton distribution functions for Tevatron and Large Hadron Collider (LHC).

Concerning lepton-nucleus colliders, there are a number of advanced projects namely JLEIC [1], eRHIC [2], LHeC [3, 4], ERL60-FCC [5] and CEPC-SppC [6]. Recently, energy frontier FCC-based $eA$ colliders are proposed in Ref. 7.

As known, China proposes a two-stage circular collider project. First stage is the Circular Electron Positron Collider (CEPC) as Higgs factory. Afterwards, SppC has been proposed as second stage dedicated for BSM new physics research, providing lead-lead opportunity as well. In addition, SppC-based $ep$ and $eA$ colliders with $E_e$ =120 GeV, $E_p$ = 37.5 TeV and $E_{Pb}$ = 3075 TeV are considered in CEPC Conceptual Design Report [6]. SppC-based energy frontier $ep$ and $\mu p$ colliders have been proposed in Ref. 8.

In this study, SppC-based lepton-nucleus collider parameters are estimated, where 0.5 TeV ILC and 5 TeV PWFA-LC electron beam energies are taken into account respectively. As to the muon beam, 1.5 TeV beam energy option of the MC is considered. Concerning their interaction point (IP) schemes, head-on collisions for all three options (i.e. ILC⊗SppC, PWFA-LC⊗SppC and MC⊗SppC), are foreseen. Schematic drawing of SppC-based lepton-nucleus colliders is shown in Fig. 1.

In Section 2, design parameters of SppC lead beam, ILC and PWFA-LC electron beams and MC muon beam are presented. Main parameters of SppC-

based "linac-ring *eA*" colliders are evaluated in Section 3. SppC-based "ring-ring *μA*" collider is considered in Section 4. Finally, results and comments are summarized in Section 5.

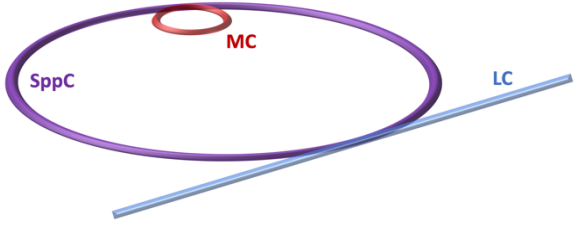

Fig. 1. SppC-based lepton-nucleus colliders

## 2. Design parameters of SppC lead beam, ILC & PWFA-LC electron beams and MC muon beam

Table 1 presents nominal parameters for 3075 TeV lead beam with 730 mA beam current. Regarding the time structure, 10080 bunches are foreseen to have 25 ns spacing under 40 MHz repetition rate.

Table 1. SppC lead beam design parameters

| Parameter [Unit] | SppC Pb beam |
|---|---|
| Beam energy, $E_{beam}$ [TeV] | 3075 |
| Number of particles per bunch | $0.18 \times 10^{10}$ |
| Normalized emittances, $\varepsilon_x^N/\varepsilon_y^N$ [μm] | 0.22 |
| $\beta_x/\beta_y$ @ IP [mm] | 750 |
| Transerve beam sizes, $\sigma_x/\sigma_y$ [nm] | 3223.8 |
| Number of bunches | 10080 |
| Repetition rate [MHz] | 40 |
| Bunch spacing [ns] | 25 |
| Bunch length, $\sigma_z$ [mm] | 70 |
| Beam current [mA] | 730 |

In Tables 2-3, 0.5 TeV and 5 TeV electron beam options of ILC and PWFA-LC are given respectively. In addition, 1.5 TeV muon beam parameters are summarized in Table 4.

Table 2. Main parameters of 0.5 TeV ILC electrons

| Parameter [Unit] | ILC electrons |
|---|---|
| Beam energy, $E_{beam}$ [TeV] | 0.5 |
| Number of particles per bunch | $1.74 \times 10^{10}$ |
| Normalized emittances, $\varepsilon_x^N/\varepsilon_y^N$ [μm] | 10/0.03 |
| $\beta_x/\beta_y$ @ IP [mm] | 11/0.23 |
| Transerve beam sizes, $\sigma_x/\sigma_y$ [nm] | 335/2.7 |
| Number of bunches | 2450 |
| Repetition rate [Hz] | 4 |
| Bunch spacing [ns] | 366 |
| Bunch length, $\sigma_z$ [mm] | 0.225 |

Table 3. Main parameters of 5 TeV PWFA-LC electrons

| Parameter [Unit] | PWFA-LC electrons |
|---|---|
| Beam energy, $E_{beam}$ [TeV] | 5 |
| Number of particles per bunch | $1 \times 10^{10}$ |
| Normalized emittances, $\varepsilon_x^N/\varepsilon_y^N$ [μm] | 10/0.035 |
| $\beta_x/\beta_y$ @ IP [mm] | 11/0.099 |
| Transerve beam sizes, $\sigma_x/\sigma_y$ [nm] | 106/0.598 |
| Number of bunches | 1 |
| Repetition rate [kHz] | 5 |
| Bunch spacing [ms] | 0.2 |
| Bunch length, $\sigma_z$ [μm] | 20 |

Table 4. Main parameters of 1.5 TeV MC muons

| Parameter [Unit] | MC muons |
|---|---|
| Beam energy, $E_{beam}$ [TeV] | 1.5 |
| Circumference [km] | 4.5 |
| Number of particles per bunch | $2 \times 10^{12}$ |
| Normalized emittances, $\varepsilon_x^N/\varepsilon_y^N$ [mm] | 0.025 |
| $\beta_x/\beta_y$ @ IP [cm] | 0.5 |
| Transerve beam sizes, $\sigma_x/\sigma_y$ [μm] | 3 |
| Number of bunches | 1 |
| Repetition rate [Hz] | 12 |
| Bunch length, $\sigma_z$ [cm] | 0.5 |

## 3. SppC-based "linac-ring *eA*" colliders

Construction of linear colliders (i.e. ILC/PWFA-LC) as well as MC tangential to SppC, will provide *eA* and *μA* collisions. In this respect, 3075 TeV lead parameters for 100 km-circumference option of the SppC, are taken into account to optimize luminosities of electron-nucleus and muon-nucleus collisions. Luminosity, beam-beam effects and disruption parameters for *e-Pb* collisions are estimated by Eqs. 1-3 respectively.

$$L_{ePb} = \frac{N_e N_{Pb}}{4\pi \sigma_{Pb}^2} f_{c_e} \quad (1)$$

$$\xi_{Pb} = \frac{N_e r_{Pb} \beta_{Pb}^*}{4\pi \gamma_{Pb} \sigma_e^2} \quad (2)$$

$$D_e = \frac{Z_{Pb} N_{Pb} r_e \sigma_{zPb}}{\gamma_e \sigma_{Pb}^2} \qquad (3)$$

As to the beam-beam effects in Eq. 2, $\gamma_{Pb} = E_{Pb}/m_{Pb}$ and $r_{Pb} = (Z_{Pb}^2/A_{Pb})r_p$. Considering linac-ring ILC⊗SppC *eA* option, 2.35x10$^{29}$ cm$^{-2}$s$^{-1}$ luminosity for 78.4 TeV center of mass energy can be achieved (see Table 5).

Table 5. ILC ⊗ SppC *eA* option

| Parameter [Unit] | ILC ⊗ SppC | |
|---|---|---|
| Beam energy, $E_{beam}$ [TeV] | e$^-$ beam: 0.5 | Pb$^{82+}$ beam: 3075 |
| Transerve beam sizes [nm] | 3223.8 | 3223.8 |
| Number of particles per bunch | 1.74x10$^{10}$ | 0.18x10$^{10}$ |
| Longitudinal beam size, $\sigma_z$ [mm] | 0.225 | 70 |
| Beam-beam effect, $\xi_{Pb}$ | - | 0.31 (0.02$^{\#}$) |
| Number of bunches | 2450 | 10080 |
| Revolution frequency, $f_{rev}$ [kHz] | - | 3 |
| Circumference, [km] | - | 100 |
| Beam current, I [mA] | 7.6 | 730 |
| Disruption, $D_e$ | 2.86 | - |
| Collider | | |
| Center of mass energy, $E_{c.m.}$ [TeV] | 78.4 | |
| Luminosity, L [cm$^{-2}$s$^{-1}$] | 2.35x10$^{29}$ | |
| Collision frequency, $f_{coll}$ [kHz] | 9.8 | |

Table 6. PWFA-LC ⊗ SppC *eA* option

| Parameter [Unit] | PWFA-LC ⊗ SppC | |
|---|---|---|
| Beam energy, $E_{beam}$ [TeV] | e$^-$ beam: 5 | Pb$^{82+}$ beam: 3075 |
| Transerve beam sizes [nm] | 3223.8 | 3223.8 |
| Number of particles per bunch | 1x10$^{10}$ | 0.18x10$^{10}$ |
| Longitudinal beam size, $\sigma_z$ [mm] | 0.02 | 70 |
| Beam-beam effect, $\xi_{Pb}$ | - | 0.18 (0.02$^{\#}$) |
| Number of bunches | 1 | 10080 |
| Revolution frequency, $f_{rev}$ [kHz] | - | 3 |
| Circumference, [km] | - | 100 |
| Beam current, I [mA] | 0.008 | 730 |
| Disruption, $D_e$ | 0.28 | - |
| Collider | | |
| Center of mass energy, $E_{c.m.}$ [TeV] | 248 | |
| Luminosity, L [cm$^{-2}$s$^{-1}$] | 6.89x10$^{28}$ | |
| Collision frequency, $f_{coll}$ [kHz] | 5 | |

Considering upgraded $\xi^{\#}_{Pb}$ in Tables 5 and 6, 0.02 value is plausible for single IP by moderate decrement of $\beta_{Pb}$. This can technically be achieved by increase of emittance at IP, while keeping the transverse beam sizes fixed, resulting in steady luminosities for both ILC⊗SppC and PWFA-LC⊗SppC options.

In Table 6, it is pointed out that 6.89x10$^{28}$ cm$^{-2}$s$^{-1}$ luminosity for 248 TeV center of mass energy is feasible via linac-ring PWFA-LC⊗SppC *eA* option.

Concerning luminosity, beam-beam effects and disruption estimations in Tables 5-7, interacting round beams are transversely matched to each other. One can see that the disruption parameters are pretty plausible for both ILC⊗SppC and PWFA-LC⊗SppC options.

**4. SppC-based "ring-ring *μA*" collider**

Regarding ring-ring *μA* collider, luminosity and beam-beam effects are estimated by Eqs. 4-6 respectively.

$$L_{\mu Pb} = \left(\frac{N_{Pb}}{N_\mu}\right)\left(\frac{\sigma_\mu}{max[\sigma_{Pb},\sigma_\mu]}\right)^2 L_{\mu\mu} \qquad (4)$$

$$\xi_\mu = \frac{Z_{Pb} N_{Pb} r_\mu \beta_\mu^*}{4\pi\gamma_\mu \sigma_{Pb}^2} \qquad (5)$$

$$\xi_{Pb} = \frac{N_\mu r_{Pb} \beta_{Pb}^*}{4\pi\gamma_{Pb} \sigma_\mu^2} \qquad (6)$$

In Table 7, main parameters of the MC⊗SppC are summarized. 1.82x10$^{32}$ cm$^{-2}$s$^{-1}$ luminosity for 135.8 TeV center of mass energy is achievable via ring-ring MC⊗SppC *μA* option.

As to the beam-beam effects of MC⊗SppC, it is seen that nominal parameters induce $\xi_{Pb}$ values more than it should be. Reducing $\xi_{Pb}$ is no doubt decrement of leptons per bunch, resulting in corresponding decrease of luminosity.

Table 7. MC ⊗ SppC μA option

| Parameter [Unit] | MC ⊗ SppC | |
|---|---|---|
| Beam energy, $E_{beam}$ [TeV] | μ beam: 1.5 | $Pb^{82+}$ beam: 3075 |
| Transverse beam sizes [nm] | 3223.8 | 3223.8 |
| Number of particles per bunch | $2 \times 10^{12}$ | $0.18 \times 10^{10}$ |
| Longitudinal beam size, $\sigma_z$ [mm] | 5 | 70 |
| Beam-beam effects, $\xi_\mu / \xi_{Pb}$ | 0.005 | 36.01 |
| Number of bunches | 1 | 10080 |
| Revolution frequency, $f_{rev}$ [kHz] | 66 | 3 |
| Circumference, [km] | 4.5 | 100 |
| Beam current, I [mA] | 21.1 | 730 |
| Collider | | |
| Center of mass energy, $E_{c.m.}$ [TeV] | 135.8 | |
| Luminosity, L [$cm^{-2}s^{-1}$] | $1.82 \times 10^{32}$ | |
| Collision frequency, $f_{coll}$ [kHz] | 66 | |

## 5. Conclusion

In this study, SppC's 3075 TeV lead beams are considered to collide with electrons of ILC & PWFA-LC as well as with muons of MC. Luminosity, beam-beam effects and disruption parameters are evaluated. It is shown that luminosities of order of $10^{30}$ $cm^{-2}s^{-1}$ can be achieved by moderate upgrades of lepton and nucleus beam parameters.

Construction of linear colliders (i.e. ILC and PWFA-LC) tangential to SppC will enhance physics search potential for SM as well as BSM phenomena. Hence, state-of-the-art lepton-nucleus colliders such as: ILC⊗SppC, PWFA-LC⊗SppC and MC⊗SppC, are indispensable for planning the future of high energy physics.

Since PWFA-LC⊗SppC and MC⊗SppC options need to be developed, feasibility of ILC⊗SppC seems a bit more realistic for *eA* collisions. On the other hand, a standalone electron linac or a standalone muon ring tangential to SppC undoubtedly requires dedicated design efforts.

**Acknowledgement**

The author would like to thank Prof. Saleh Sultansoy for useful discussions.